\documentclass[9pt,twocolumn,twoside]{opticajnl}
\journal{opticajournal} 

\setboolean{shortarticle}{false}

\usepackage{siunitx}
\usepackage{lineno}
\usepackage{stfloats}
\usepackage{comment}
\usepackage{array}
\linenumbers 

\title{Cryogenic Optical Packaging Using Photonic Wire Bonds}

\author[1,2]{Becky Lin}
\author[1,2]{Donald Witt}
\author[2,3]{Jeff F. Young}
\author[1,2]{Lukas Chrostowski}

\affil[1]{Department of Electrical and Computer Engineering, The University of British Columbia, V6T 1Z4, Vancouver, British Columbia, Canada}
\affil[2]{Stewart Blusson Quantum Matter Institute, V6T 1Z4, Vancouver, British Columbia, Canada}
\affil[3]{Department of Physics and Astronomy, The University of British Columbia, V6T 1Z1, Vancouver, British Columbia, Canada}

\affil[*]{becky.lin@alumni.ubc.ca, lukasc@ece.ubc.ca}

\begin{abstract}
We present the required techniques for the successful low loss packaging of integrated photonic devices capable of operating down to 970 mK utilizing photonic wire bonds. This scalable technique is shown to have an insertion loss of less than 2 dB per connection between a SMF-28 single mode fibre and a silicon photonic chip at these temperatures. This technique has shown robustness to thermal cycling and is ultra-high vacuum compatible without the need for any active alignment.
\end{abstract}

\setboolean{displaycopyright}{false} 

\begin{document}
\nolinenumbers
\maketitle

\section{Introduction}
In this era of emerging quantum technology, there is a growing demand for easy and efficient optical connection methodologies for cryogenic photonic circuits. Low temperature operation suppresses thermal noise and makes available special material properties, such as superconductivity, for single photon detectors \cite{akhlaghi2015waveguide, ferrari2018waveguide,detectorphotoncounting}, and single photon sources \cite{integratedQD}. Efficient cryogenic optical coupling is needed for practical implementations for all-optical quantum computing \cite{o2007optical, rudolph2017optimistic}, hybrid quantum photonic integrated circuits \cite{yan2021silicon, moody20222022, elshaari2020hybrid}, linking together nodes of the quantum internet \cite{quantumnetwork}, and quantum communication \cite{qkdreceiver}. Low insertion loss (IL) is especially important for photonic connections that transport quantum information in the form of single photons as photon loss can result in reduced photon pair generation rate \cite{afifi2021contra, silverstone2014chip} and increased overhead for error correction \cite{yan2020quantum}.

\begin{figure}[h!]
\centering\includegraphics[width=0.45\textwidth]{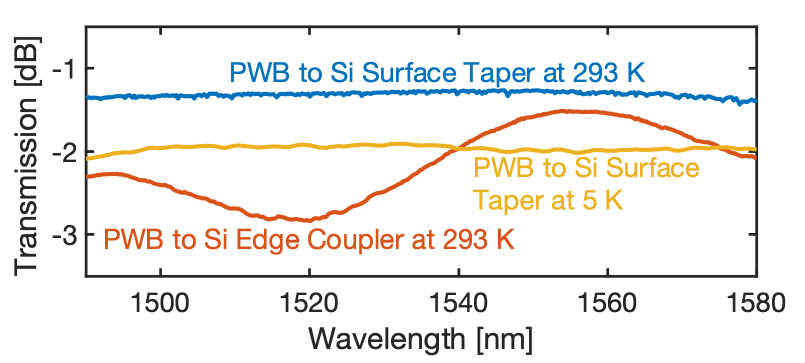}
\caption{Plot showing the transmission of best achieved in-house bonded photonic wire bonds (PWBs) connecting between a fiber to silicon photonic waveguide via surface tapers (blue) and edge couplers (red) at room temperature. The transmission is calibrated for one PWB. Measurements are optimized to TE polarization at \SI{1550}{\nano\meter} using a polarization paddle. TE calibrated transmission from fiber to surface taper through one PWB at \SI{5}{\kelvin} demonstrated in this work is shown in yellow. The IL at \SI{5}{\kelvin} for C band is estimated to be 2.0 \(\pm\) \SI{0.3}{\decibel}.}
\label{fig:fig1}
\noindent\makebox[\linewidth]{\rule{\linewidth}{0.4pt}}
\end{figure}

\begin{figure*}[bp!]
\centering
\makebox[\textwidth]{\includegraphics[width=170mm]{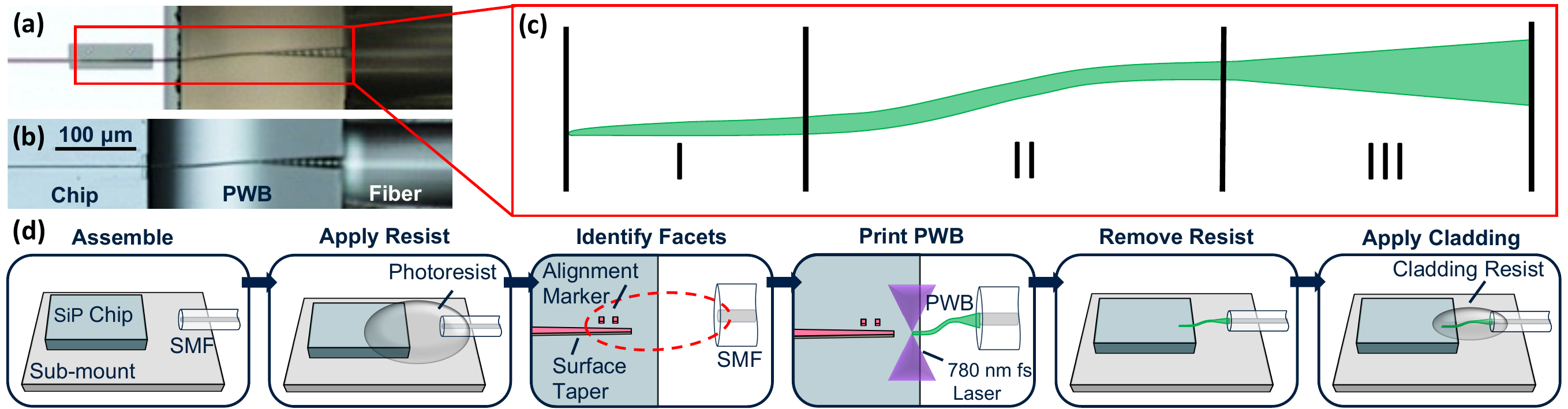}}
\caption{(a) Adiabatic coupling via surface taper. The box around the surface taper is etched silicon from a positive electron beam lithography process. The white background is unetched silicon. (b) Mode matched coupling via edge coupler. (c) The PWB consist of three sections: taper to evanescently couple light to surface taper (I), uniform cross section waveguide connecting section I and III (II), and a taper to adiabatically expand the mode to match to the fiber core (III). (d) Generic workflow to assemble and connect optical fiber to waveguide using a photonic wirebond written using two photon polarization.}
\label{fig:fig2}
\end{figure*}

In this paper, we use photonic wire bonding (PWB) to make reliable and efficient optical connections between single mode fiber (SMF) and silicon photonic circuits down to \SI{970}{\milli\kelvin}. PWB is a new emerging optical integration technique that uses a polymer based free form structure to connect two roughly aligned (> \SI{30}{\micro\meter} of x-y tolerance, > \SI{100}{\micro\meter} of tolerance along optical axis) optical ports \cite{lindenmann2018photonic}. This 3D optical bridge eliminates the precise alignment required for other integration methods. The ability to write adiabatic mode conversion tapers allows for efficient coupling between the large area fiber mode (mode field diameter \(\sim\) \SI{10.4}{\micro\meter}) to the small area waveguide mode (mode field diameter \(\sim\) \SI{0.5}{\micro\meter}). The resist (VanCore A) of the bond has low absorption in the O and C bands, allowing for a large working bandwidth (see Supplement 1 for measurement of resist transmission). These properties are attractive for quantum applications that require the connection of color centers with different operating wavelengths, allowing for the creation of hybrid integrated quantum photonic circuits \cite{elshaari2020hybrid}. Its freeform nature also allows for maintenance and manipulation of polarization \cite{nesic2021ultra}. Versatility has been demonstrated at room temperature by the creation of fibre-to-chip \cite{lindenmann2014connecting}, laser-to-chip \cite{billah2018hybrid}, and chip-to-chip \cite{blaicher2020hybrid,xu2021hybrid} interfaces. Fig.~\ref{fig:fig1} shows the best achieved IL of in-house bonded fiber to surface taper (blue) [Fig.~\ref{fig:fig2}(a)] and fiber to edge coupler (red) [Fig.~\ref{fig:fig2}(b)] at room temperature (RT), which are \SI{1.3}{\decibel} and \SI{1.6}{\decibel} at \SI{1550}{\nano\meter}, respectively. The on-chip surface tapers and edge couplers can be formed using standard fabrication steps offered by silicon photonic foundries. Unlike other adiabatic tapers that require coupling to specially made tapered fibers \cite{burek2017fiber}, PWBs can directly bond to commercially available cleaved fibers and fiber arrays (FA). PWB samples are also mechanically robust which makes them more immune to vibrations from nearby pumps and compressors.

The main challenges of optical coupling between a fibre and a integrated photonic circuit inside a cryogenic chamber stem from: (1) thermal induced stresses and strains (2) inaccessibility of the closed testing setup, and (3) the limitation of available cryogenic compatible materials. Thus, it can not be assumed a device demonstrated to have good performance at room temperature will remain so at low temperature. Thermal contraction during cool down can introduce relative displacements that can jeopardize pre-aligned systems or break pre-packaged samples \cite{you2020chip}. However, without the camera and hardware manipulators usually available for room temperature setups, it is difficult to access and adjust samples that are loaded and sealed inside cryogenic chambers \cite{macdonald2015optical}.

In this context, the PWB technique offers some significant advantages. First, the glue used no longer needs to be optically matched to, or physically in contact with the coupling interface. This significantly reduces the constraint on the glue used since optically clear glues, which are often UV activated, are prone to crack at low temperatures \cite{wasserman2022cryogenic}. It also avoids the problem of spectral shifts caused by the glue’s changing refractive index during cool down \cite{mckenna2019cryogenic}. Second, no active and precise alignment is needed during assembly \cite{zeng2023cryogenic} as this is achieved through image recognition by the PWB tool during the bonding process. Third, assembly time and valuable real estate are saved when multiple channels are assembled at once using a FA. By using FAs with up to 64 channels on a \SI{127}{\micro\meter} pitch, access to other important surface elements such as electronic pads can be greatly improved.

In this paper, we show the assembly methodology for cryogenic optical packaging using PWB and demonstrate coupling of light between fibers to surface tapers down to \SI{970}{\milli\kelvin} with an IL of \SI{2.0}{\decibel} per coupler at \SI{1550}{\nano\meter}. We evaluate the performance of PWBs in ultra high vacuum (UHV) environment and various input power levels. We use a micro ring resonator (MRR) to demonstrate that we can successfully couple sufficient optical power to observe non-linear behaviour at low temperatures.

\begin{figure*}[h!]
\centering
\includegraphics[width=170mm]{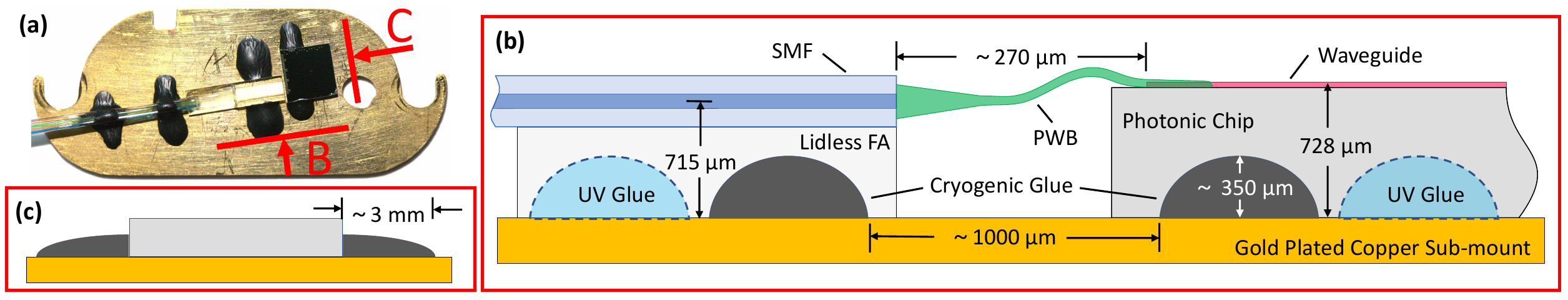}
\caption{(a) Image of successfully cooled sample. Roughly equal amounts of cryogenic glue (Masterbond® EP29LPSPAO-1) are applied to opposite sides of chip and FA to help equalize thermal induced stresses. The amount of glue is minimized to reduce stresses inside the glue and between the glue and the other components. (b) Side schematic view of working sample. The UV glue is used to temporarily hold the chip and FA to the sub-mount while the cryogenic glue cured over 12 hours and is removed cleanly by a scalpel afterwards. The core of the SMF sits slightly (10 \(\sim\) \SI{20}{\micro\meter}) below the silicon waveguide. The distance between the SMF face and silicon waveguide taper tip is \(\sim\) \SI{270}{\micro\meter} to allow room for mode transitioning taper and relaxed bond arc. (c) Both the cryogenic and temporary UV glue are applied to the sides of the FA and chip only and not underneath them.}
\label{fig:fig3}
\noindent\makebox[\linewidth]{\rule{\textwidth}{0.4pt}}
\end{figure*}

\section{Background}
A PWB consists of three sections. Namely, a polymer taper on one end to evanescently couple the mode to the tapered silicon waveguide (I in Fig.~\ref{fig:fig2}(c)), a polymer taper on the opposite end to adiabatically transition the mode the fiber core (III in Fig.~\ref{fig:fig2}(c)) and a uniform polymer waveguide connecting the two tapers together with a free form path (II in Fig.~\ref{fig:fig2}(c)). The polymer resist has a refractive index of \(\sim\) 1.531 at \SI{1550}{\nano\meter}. The silicon waveguide has a cross-section design width of \SI{500}{\nano\meter} and height of \SI{220}{\nano\meter} while the surface taper has a design length of \SI{65}{\micro\meter} and tip width of \SI{130}{\nano\meter}. The waveguides are not cladded and are thus exposed to the environment.

An assembled sample consists of the photonic chip and fiber, or FA, glued onto a common sub-mount. As photonic wire bonding allows for high alignment tolerance, the assembly can be made by hand using tweezers under a stereomicroscope. An overview of the generic photonic wire bonding process is shown in Fig.~\ref{fig:fig2}(d). Namely, it consists of assembling the sample, applying the polymer resist to the area of interest, using the Vanguard SONATA1000 tool to identify the location of facets and write the bonds using the two-photon polymerization process, developing the resist using solvents, and protecting the PWB with a cladding resist (cladding is not applied in this work, see section 3.\ref{Cladding} for details).

\section{Cryogenic Photonic Wire Bonding Process}

\subsection{Sample Assembly}
The procedure for cryogenic photonic wire bonding adapts the room temperature PWB process but with special considerations for successful cryogenic operation. An assembled sample consists of a silicon photonic chip and an 8-channel single mode FA mounted on a gold-plated oxygen-free copper sub-mount. Copper is chosen for its high thermal conductivity and the gold plating maintains good thermal contact between the sub-mount and the gold-plated cold stage of the cryostat \cite{ekin2006experimental}. The jacket of the FA is a thermal plastic (Hytrel) that does not effect the vacuum. A cryogenic (and vacuum) compatible two-part epoxy (Masterbond® EP29LPSPAO-1 Black \cite{EP29LPSPAOref}) is used instead of the UV glue that is normally used to assemble room temperature devices. 

Fig.~\ref{fig:fig3}(a) shows an image of a working sample and Fig.~\ref{fig:fig3}(b) shows its side schematic view. The face of the FA is put  \(\sim\) \SI{270}{\micro\meter} from the tip of the surface taper. The heights of the fiber core and the silicon waveguide are \SI{715}{\micro\meter} and \SI{728}{\micro\meter}, respectively. This sets the fiber core slightly below the waveguide. The core height is standard for a \SI{750}{\micro\meter} FA v-groove. The glue anchors for the FA and the chip are roughly \SI{1000}{\micro\meter} apart. The volume and position of the glue is roughly mirrored on both sides of the FA and chip to help equalize any thermally induced stress. No glue is applied underneath the FA or chip [Fig.~\ref{fig:fig3}(c)]. Glue is also applied to the fiber ribbon as strain relief.

The samples assembled using the configuration described above did not break when cooled to cryogenic temperatures. See Supplement 1 for the type of assemblies that do not work.

\subsection{Thermal Contraction Matching and Stress Management}

The large amount of thermal contraction the sample undergoes while cooling to cryogenic temperatures necessitates careful design considerations. Sub-components of the sample, comprising the silicon photonic chip, copper sub-mount, and polymer-based PWB, undergo different rates of thermal contraction during cool down. Polymer has the highest thermal contraction (\(\Delta\)L/L = \SI{1.22}{\percent} from \SI{293}{\kelvin} to \SI{4}{\kelvin}) followed by copper (\(\Delta\)L/L = \SI{0.324}{\percent} from \SI{293}{\kelvin} to \SI{4}{\kelvin}) and silicon (\(\Delta\)L/L = \SI{0.022}{\percent} from \SI{293}{\kelvin} to \SI{4}{\kelvin}) \cite{ekin2006experimental,reed1983materials}. Here, we assume the PWB polymer is acrylate based and has similar properties as methyl methacrylate (PMMA) \cite{o2023two} and use total thermal linear contraction values (\(\Delta\)L/L \(\equiv\) (\(L_\text{\SI{293}{\kelvin}}\)-\(L_{T}\))/\(L_\text{\SI{293}{\kelvin}}\)) measured for PMMA as a reference for the PWB.

The contraction of the bond can be roughly matched to the contraction of the copper sub-mount by adding a gap between the chip and FA. The estimated contraction for a \SI{270}{\micro\meter} polymer bond is \SI{3.3}{\micro\meter} which will require a copper length of around \SI{1020}{\micro\meter} to match. This distance is set by choosing the corresponding anchoring points of the glue between the chip and FA [Fig.~\ref{fig:fig3}(b)].



\subsection{Cladding} \label{Cladding}

The cladding (n \(\approx 1.393\) at \SI{1550}{\nano\meter}) normally used to protect the PWB and reduce the insertion loss by \(\sim\) \SI{0.2}{\decibel} at room temperature is not applied to cryogenic samples, as it would out-gas in the cryostat. Attempts to mitigate the out-gassing by encapsulating the cladding resist with UHV compatible epoxy failed, likely due to stresses caused by the differing coefficients of expansion of the additional cladding materials. 

\section{Experiment and Results}
\subsection{Experimental Setup}

The photonically wire bonded devices were cooled to \SI{5}{\kelvin} inside a custom closed cycle, UHV dry cryostat. Samples are cooled from \SI{300}{\kelvin} to \SI{5}{\kelvin} within \(\sim\) 4 hours and remain at \SI{5}{\kelvin} for > 17 hour before being warmed back up to \SI{300}{\kelvin}. For both baking and repeatability cycle tests, the sample is left inside the vacuum environment of cryostat. The devices were also tested in a different custom  dry cryostat (non-UHV) capable of cooling to \SI{970}{\milli\kelvin} to test their performance at even lower temperature.  

\begin{figure}[h!]
\noindent\makebox[\linewidth]{\rule{\linewidth}{0.4pt}}
\centering\includegraphics[width=0.45\textwidth]{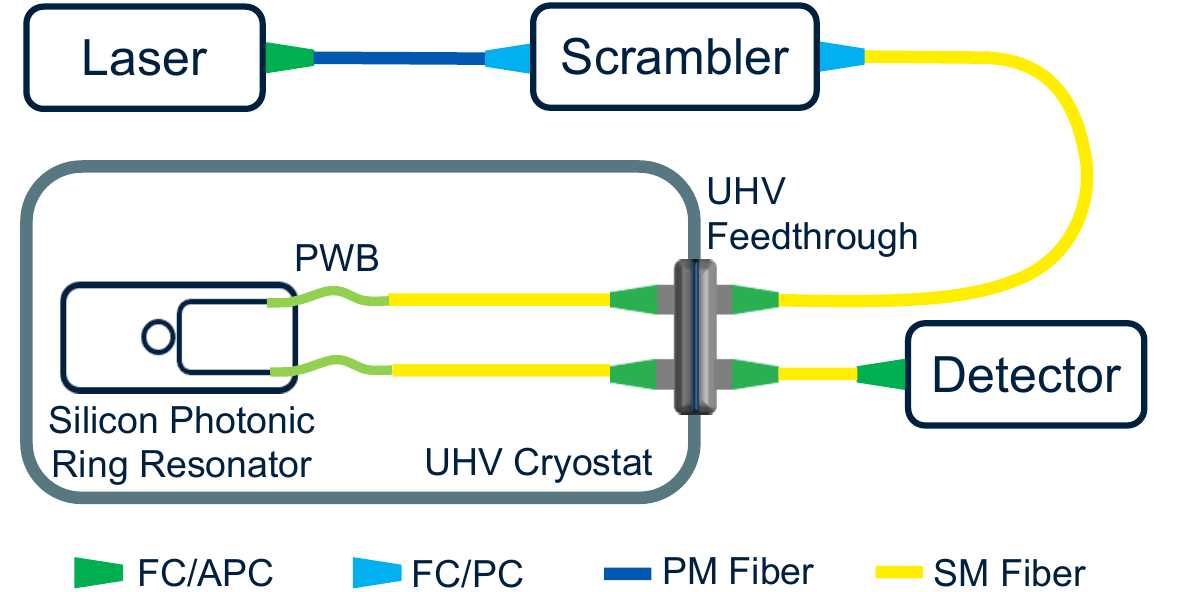}
\caption{Optical pathway of measurement.}
\label{fig:fig4}
\noindent\makebox[\linewidth]{\rule{\linewidth}{0.4pt}}
\end{figure}

The full optical path of the measurement is shown in Fig.~\ref{fig:fig4}. The devices are optically measured using a Keysight 8164B mainframe equipped with an 81682A tunable laser module and 81635A power sensor. The light passes from the tunable laser module into a PS-155-A FiberPro polarization scrambler. The light then passes into the cryostat through a UHV FC/APC fiber feedthrough from SQS Vlaknova Optika. Then the light passes through the sample and out a different port on the same feedthrough and into the 81635A power sensor.

The device is continuously measured over the full laser range during the whole process to monitor the effect of each step. 

\subsection{Optical Performance of PWB}
\begin{figure*}[bp!]
\centering
\makebox[\textwidth]{\includegraphics[width=170mm]{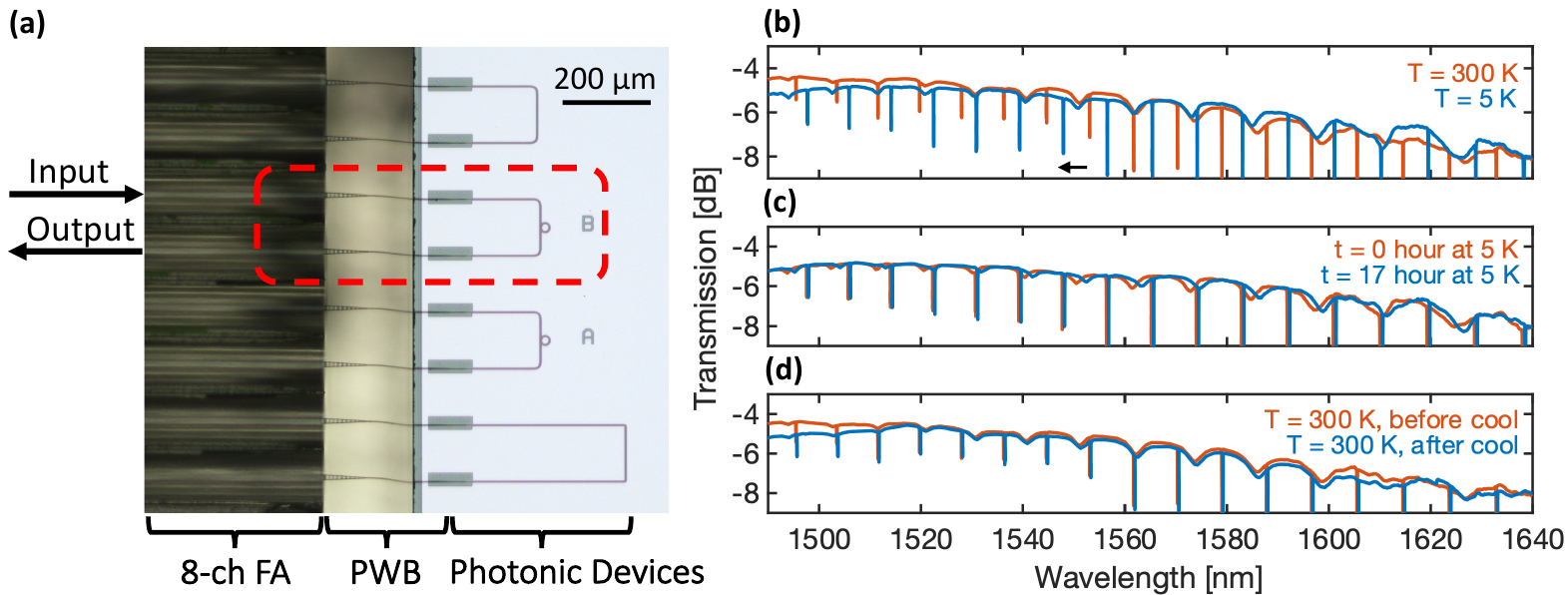}}
\caption{(a) Optical input from one fiber is guided to the surface taper through the PWB and coupled into the MRR. Light is coupled out through the other fiber port. The cool down and warm up measurement for MRR B (circled in red) is shown on the right: the calibrated transmission of two PWBs connected to a MRR (b) before cool down at \SI{300}{\kelvin} vs after cool down at \SI{5}{\kelvin} (c) before vs after maintaining at \SI{5}{\kelvin} for 17 hours (d) before cooling down at \SI{300}{\kelvin} vs after warming from \SI{5}{\kelvin} at \SI{300}{\kelvin}. The change in the temperature of the ring is evident from the shift in resonance wavelength. The transmission remains approximately the same through the cool down and warm up process.}
\label{fig:fig5}
\end{figure*}

To measure the IL of the PWB and to demonstrate the coupling of light to photonic devices at cryogenic temperatures, waveguide loopbacks and MRRs structures are used [Fig.~\ref{fig:fig5}(a)]. As only one device can be measured during each cooling cycle, the MRR device (circled in red) is prioritized as it can provide information about the temperature and surface condition of the device for comparison to the temperature and pressure sensors inside the cryostat. For a consistent measurement, the system tracked the same device over the cycle of cool down and warm up to examine the performance over temperature and time.

The plots in Fig.~\ref{fig:fig5} (b)-(d) show the measured transmission through the two PWBs and MRR device before and after cooling. Prior to cooling, at \SI{300}{\kelvin}, the transmission ranges from \SI{-4.5}{\decibel} at \SI{1490}{\nano\meter} to \SI{-8.0}{\decibel} at \SI{1640}{\nano\meter}. After cooling to \SI{5}{\kelvin}, the transmission at \SI{1490}{\nano\meter} is \SI{-5.2}{\decibel} and \SI{-8.0}{\decibel} at \SI{1640}{\nano\meter} [Fig.~\ref{fig:fig5}(b)]. Specifically, the transmission in the C-band is \num{-5.3} \(\pm\) \SI{0.2}{\decibel} which is relatively wavelength independent compared to methods that use grating couplers. 

The sample was left at \SI{5}{\kelvin} for 17 hours to study the stability. Figure \ref{fig:fig5}(c) shows the transmission spectrum before and after cooling for 17 hours. There is no change in the IL and the slight red shift in the resonance is due to the deposition of condensates on the ring over time in a unbaked vacuum system. No sign of degradation in the PWBs is observed. Figure \ref{fig:fig5}(d) compares the sample prior to cooling at \SI{300}{\kelvin} to after warming up (from \SI{5}{\kelvin}) at \SI{300}{\kelvin}. Overall, there is a slight increase of IL by less than \SI{0.2}{\decibel}.

The IL of a single PWB is approximated by dividing the calibrated transmission through two bonds by two. We assume the loss through the short silicon waveguide (\SI{430}{\micro\meter}) and the two bends is negligible. The resulting IL of the PWB at \SI{5}{\kelvin} measured using the polarization scrambler is then \num{2.7} \(\pm\) \SI{0.1}{\decibel} at \SI{1550}{\nano\meter}. Accounting for the \num{0.7} \(\pm\) \SI{0.2}{\decibel} of polarization dependent loss if the input is optimized for TE mode (typically done for wavelength specific measurements), the IL of the PWB would be \num{2.0} \(\pm\) \SI{0.3}{\decibel} (yellow in Fig.~\ref{fig:fig1}). The calibration procedure is described in more detail in Supplement 1.

The periodic shallow dips that are present in addition to the sharp resonance peaks are likely the resonances of the TM mode introduced by the polarization scrambler. The quality factor is expected to be lower for the TM mode resonances as the coupling region of the ring was designed for TE coupling. The TM resonances are more prominent at longer wavelengths because the mode at shorter wavelengths is more contained inside the waveguide and is minimally coupled to the ring.

The sample was also cooled to \SI{970}{\milli\kelvin} using a separate (non-UHV) dry cryostat.  It survived temperature cylcing and performed the same at \SI{970}{\milli\kelvin} as it did at \SI{5}{\kelvin}.

\begin{figure*}[h!]
\centering
\makebox[\textwidth]{\includegraphics[width=170mm]{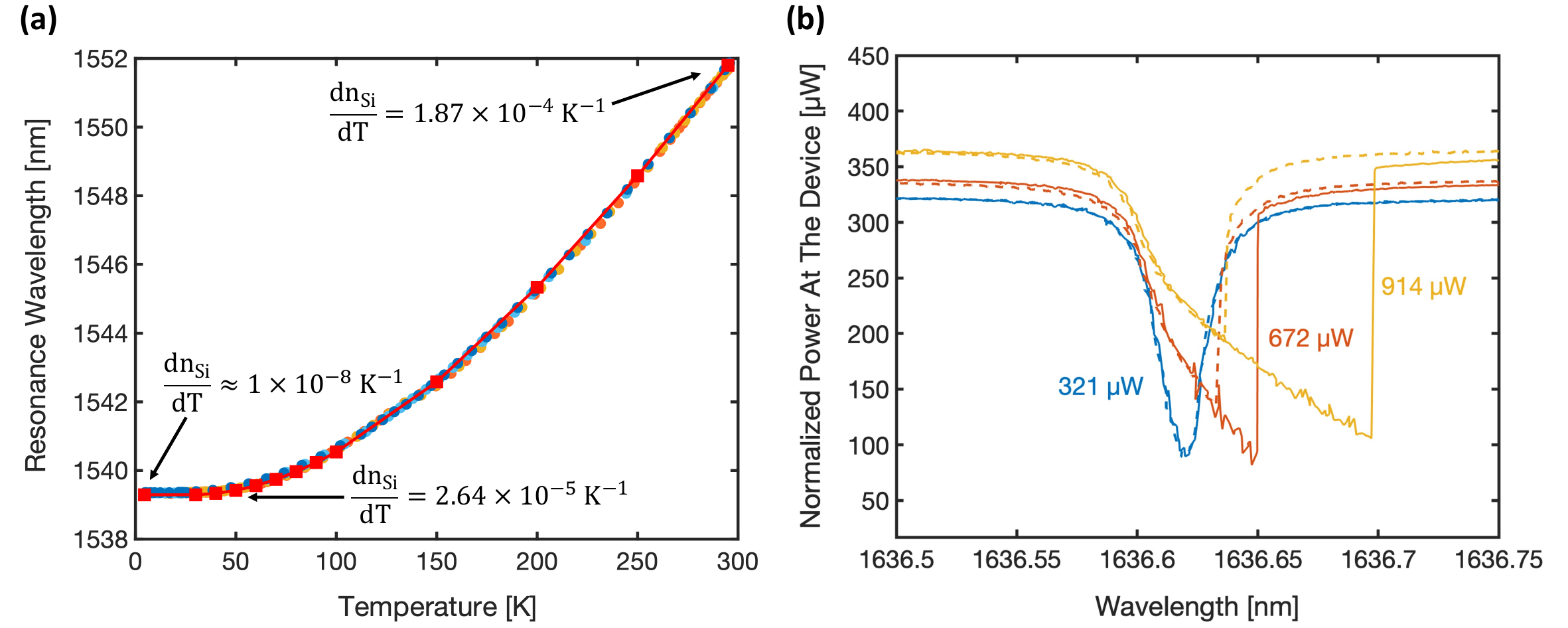}}
\caption{(a) The system is baked at \SI{120}{\degreeCelsius} to reach UHV pressure of \(\sim 9.2 \times 10^{-9}\)  mbar. A plot of resonance wavelength versus temperature shows that the resonance remains the same over multiple cool down and warm up cycle and agrees well with simulation (red). The simulation curve is shifted by \SI{0.1}{\nano\meter} to match to the measured wavelength at \SI{295}{\kelvin}. The change in resonance wavelength is dominated by the thermal optic effect of the silicon waveguide which is 22 times larger than the buried oxide. (b) The transmission spectra of a silicon photonic MRR operating at \SI{5}{\kelvin} for different input powers. The blue curve is used as a reference with no nonlinearity visible. The other curves have been normalized by subtracting the laser input power setting so that they overlapped for direct comparison. The input power at the device is noted on the plot. The dashed lines indicate the response generated by scanning the laser in the reverse direction. One can see the difference between the forward and reverse scans for the red and the yellow spectra.}
\label{fig:fig6}
\noindent\makebox[\linewidth]{\rule{\textwidth}{0.4pt}}
\end{figure*}

\subsection{Multiple Cool Down Performance}
The sample underwent three cool down cycles to test the performance repeatability and the behaviour remained the same. Namely, the IL is observed to degrade slightly during cool down, remain the same for several hours at \SI{5}{\kelvin}, and return to the original performance after warming up. Overall, it is observed a working sample can withstand the multiple cycles of cooling and long cooling times over the several weeks of operation. See the Supplement 1 for additional data in support of these claims.  

\subsection{Operation in Vacuum Conditions}
\label{vacuum_section}
Supplement 1 describes how gas adsorption gradually modifies the resonant frequency of the MRR devices \cite{gastunning1,gastunning2} held at \SI{5}{\kelvin} for several hours.  We therefore also tested the ability of the sample to survive vacuum bake out prior to cooling. The system, which includes the cryogenic chamber and sample, was heated to \SI{120}{\degreeCelsius} for 166 hours to achieve a final pressure of \(\sim 9.2 \times 10^{-9}\) mbar. By achieving UHV conditions before cooling to cryogenic temperatures, the effects of undesirable gas molecule deposition were negligible over the course of our experiments which acquired data over several days (including that shown in Fig. ~\ref{fig:fig6}(a)). However, the baking procedure increased the IL by approximately \SI{3}{\decibel} (\SI{1.5}{\decibel} for each bond). 

\subsection{Power Limit at Different Pressures}
Several potential applications for cryogenic photonic wire bonds depend on nonlinear optical effects. These applications include photon pair sources \cite{photonpair} and all optical switching \cite{allopticalswitch}. In order to achieve these nonlinear optical effects, one needs to have high optical power in the device. To examine the suitability of photonic wire bonds in such cases, a series of power sweeps were run to determine the damage threshold of the assemblies.

In normal atmospheric conditions, the power limit of the photonic wire bonds exceeds the \SI{13.5}{\decibel}m maximum output power of our laser. However, in a vacuum environment, whether the temperature of the sample is \SI{300}{\kelvin} or \SI{5}{\kelvin}, the threshold power  was found to be less than \SI{6}{\decibel}m when continuously sweeping the laser wavelength over the standard scan range.

Optical images of the sample taken after such high power sweeps show the middle section of the photonic wire bonds has evaporated (see Supplement 1). The fact that the upper power limit of a photonic wire bond is the same whether the sample is at \SI{300}{\kelvin} or \SI{5}{\kelvin} indicates that the likely cause is the lack of convection cooling in a vacuum environment. This suggests the power limit may be higher in vapour or liquid cryostats that can offer higher convection cooling for the insulating polymer bonds.

\subsection{Temperature Dependence of MRR Resonances}

We track one of the resonances of the MRR over the \SI{300}{\kelvin} - \SI{5}{\kelvin} temperature range and compare it to simulations to help verify the sample does in fact thermalize with the cold stage. Fig.~\ref{fig:fig6}(a) shows a plot of resonance wavelength versus temperature in a UHV environment. The resonance wavelength shift from \SI{295}{\kelvin} or \SI{5}{\kelvin} over several cool down (blue coloured dots) and warm up (orange coloured dots) cycles match well with the simulation (red line). The simulation is produced using refractive index values for silicon and silicon dioxide at different temperatures reported in \cite{frey2006temperature,leviton2006temperature,komma2012thermo}. See Supplement 1 for more details of the simulation method.

\subsection{Cryogenic Ring Non-linearity}

We also use the MRR to demonstrate packaging of a non-linear optical device operating at cryogenic temperatures using PWB.

The nonlinear response was found to be significantly weaker at cryogenic temperatures, as expected given the dramatically lower thermo optic coefficient of silicon at low temperatures \cite{frey2006temperature}. However, it was also found that within the acceptable power limits of the photonic wire bonds, nonlinear optical effects were still visible. An example of such a nonlinear response can be seen in Fig.~\ref{fig:fig6}(b). The solid line indicates a forward sweep of the measurement laser (from low to high wavelength). The dashed line indicates a reverse sweep of the laser (from high to low wavelength). Though much reduced, the thermo-optic effect may still be contributing to the observed redshift of the resonances (in addition to the Kerr effect) at these low temperatures.  A nonlinear model will be used to quantify the relative contributions of Kerr, thermo-optic, and free carrier effects on these low temperature, continuous wave nonlinear results. Future pulsed laser excitation experiments will also aid in separating the different nonlinear effects via differences in their time constants \cite{nonlinearyale}.

\section{Summary}
The results presented here show that PWBs can be used for photonic packaging from room temperature down to \SI{970}{\milli\kelvin} with low loss, high bandwidth, and mechanically robust performance. This method alleviates the strict alignment challenge that occurs during assembling and cool down experienced by other packaging methods. Additional features such as UHV, vacuum bakeout compatibility and robustness to thermal cycling make PWB a promising new photonic packaging technique for a wide range of applications and operating conditions.

\section{Discussion}

\sisetup{propagate-math-font = true, reset-math-version = false}

\label{comparison_supplement}
\begin{table*}[bp!]
\centering
\caption{\bf Comparison of Different Fixed Coupling Methodologies Used for Cryogenic Silicon Photonic Integration At C-Band.}

\begin{tabular}{>{\raggedright}p{3cm} >{\raggedright}p{3cm} >{\raggedright}p{3cm} >{\raggedright}p{3cm} p{3cm}}

\hline
& Tapered Fiber Glued to Waveguide \cite{wasserman2022cryogenic} & Angle Polished Fiber Glued to Grating Coupler \cite{wasserman2022cryogenic, mckenna2019cryogenic} & Fiber Array Coupled to Edge Coupler \cite{hiraki2013cryogenic} & \textbf{This Work} \\

\hline
Demonstrated Temperature   & \SI{7}{\milli\kelvin} & \SI{7}{\milli\kelvin} & \SI{4}{\kelvin} & \textbf{\boldmath {970} \unit{\milli\kelvin}} \\

Coupling Efficiency Per Coupler  & \SI{-5.2}{\decibel} & \num{-6} \(\sim\) \SI{-10}{\decibel} & \SI{-1.9}{\decibel} & \textbf{\boldmath{-2.7} \unit{\decibel} (scrambled) \newline {-2.0} \unit{\decibel} at {1550} \unit{\nano\meter} (TE optimized)}\\

Coupling Bandwidth & > \SI{100}{\nano\meter} & 20 \(\sim\) \SI{55}{\nano\meter} & > \SI{20}{\nano\meter} & \textbf{> \boldmath{100} \unit{\nano\meter}}\\

Spectral Shift & Minimal & Large & Minimal & \textbf{Minimal} \\

Polarization Dependent Loss & Minimal & Large & Minimal & \textbf{Minimal} \\

Misalignment Tolerance & High:\(\sim\) \SI{50}{\micro\meter} along optical axis & Medium:\(\sim\) \SI{2}{\micro\meter} & Low: \(\sim\) \SI{0.5}{\micro\meter} & \textbf{High: > \boldmath{30} \unit{\micro\meter} in all axis} \\

Alignment Method & Active & Active & Active & \textbf{Passive \& Image Recognition} \\

Scalability & Single Fiber & Single Fiber & Medium to High & \textbf{Medium to High}\\

\hline
\end{tabular}
  \label{tab:tab1}
\end{table*}

\subsection{Comparison of Cryogenic Coupling Techniques}

We compare the method shown in this work to other cryogenic fixed coupling methods that have been demonstrated for silicon photonics cooled to < \SI{10}{\kelvin}, which is the temperature required for many superconducting samples [Table. \ref{tab:tab1}]. This work demonstrates one of the lowest IL of these methods at \SI{2.0}{\decibel} at \SI{1550}{\nano\meter} which is approximately 3 to \SI{7}{\decibel}  less than the other methods. We show the largest coupling bandwidth, > \SI{100}{\nano\meter}, with potential to be larger in the O-band and near visible band. While spectral shift is a prominent issue for other methods that use grating based couplers during assembly and cool down, we do not observe any spectral shift. We also observe minimal polarization dependent loss. The free-form capability of the PWB also provides the largest misalignment tolerance of more than \SI{30}{\micro\meter} in all three axis which is significantly more than the 0.5 - \SI{2}{\micro\meter} tolerance of the other methods using grating couplers or edge couplers. As a result, the assembly process does not require any active alignment.  Precise bonding is achieved through image recognition in the PWB tool. This method can be used for large scale integration using commercially available fiber arrays.

The method that uses a fiber array glued to grating couplers is not included in the table as results have shown that configuration is prone to crack and break when cooled to low temperatures due to the thermal stress from the large quantity of glue needed to keep the FA intact \cite{you2020chip,de2020millivolt}. 

\subsection{Thermal Conductivity}
The thermal conductivity between the chip and the sub-mount is achieved through the thermal conductive glue as the assembly process shown in this work does not have any bonding agent applied underneath the chip. For chips with larger heat load, a thin layer of thermal conductive paste, glue, or solder may need to be applied between the chip and sub-mount to increase the thermal conductivity.

\subsection{Future Work and Improvements}
We anticipate this technique to allow medium to high scaling beyond the 8-channel coupling demonstrated in this work. Commercial FAs can have more than 64 channels which would be cumbersome to connect via FC/APC feedthroughs. Luckily, this overhead may be mitigated by using MTP® connectors that are compact and low loss. 

In addition to coupling to on-chip surface tapers, PWBs can also be bonded to edge couplers [Fig.~\ref{fig:fig2}(b)] which would be used for oxide cladded samples. 

Aside from the fundamental absorption losses in the polymer resist, the IL may be improved by new emerging 3D nano writing techniques that have improved writing resolution and minimum feature size \cite{he2022single,ouyang2023ultrafast}. This could reduce the losses due to physical properties such as scattering from surface roughness and back reflections from mode mismatch. The trajectory of the bond can also be improved by optimizing the curvatures to minimize bending loss \cite{nesic2022transformation}.

\subsection{Future Applications}

Beyond using PWB for fiber-to-chip coupling, it can be used for chip-to-chip integration to achieve hybrid quantum photonic integrated circuits. Although not all devices require cryogenic operation, it is shown that overall performance of the system improves if they are also cooled down \cite{ono2021si}. Surface taper-to-surface taper coupling has been shown to have reliable and repeatable IL of \SI{0.7}{\decibel} \cite{blaicher2020hybrid}.

\begin{backmatter}
\bmsection{Funding} 
Refined Manufacturing Acceleration Process (ReMAP); Canada Foundation for Innovation (CFI); B.C. Knowledge Development Fund (BCKDF); SiEPICfab Consortium; Natural Sciences and Engineering Research Council of Canada (NSERC); Mitacs Accelerate Program

\bmsection{Acknowledgments}
Thank you Matthias Lauermann and Matthew Mitchell for fruitful discussions.

\bmsection{Disclosures} BL: Dream Photonics (E), LC: Dream Photonics (I).

\bmsection{Data Availability Statement} 
Data underlying the results presented in this paper are not publicly available at this time but may be obtained from the authors upon reasonable request.

\bmsection{Supplemental document}
\label{supplement}
See Supplement 1 for supporting content.

\end{backmatter}

\bibliography{cryoPaper}

\bibliographyfullrefs{cryoPaper}

\ifthenelse{\equal{\journalref}{aop}}{%
\section*{Author Biographies}
\begingroup
\setlength\intextsep{0pt}
\begin{minipage}[t][6.3cm][t]{1.0\textwidth} 
  \begin{wrapfigure}{L}{0.25\textwidth}
    \includegraphics[width=0.25\textwidth]{john_smith.eps}
  \end{wrapfigure}
  \noindent
  {\bfseries John Smith} received his BSc (Mathematics) in 2000 from The University of Maryland. His research interests include lasers and optics.
\end{minipage}
\begin{minipage}{1.0\textwidth}
  \begin{wrapfigure}{L}{0.25\textwidth}
    \includegraphics[width=0.25\textwidth]{alice_smith.eps}
  \end{wrapfigure}
  \noindent
  {\bfseries Alice Smith} also received her BSc (Mathematics) in 2000 from The University of Maryland. Her research interests also include lasers and optics.
\end{minipage}
\endgroup
}{}

\end{document}


\maketitle

\section{FTIR Measurements}
\label{ftir_supplement}
It is desirable to measure the optical absorption of the Vancore A resist used in the PWB process. This information will allow for the determination of the suitability of PWB for coupling to various quantum emitters. Fourier-transform infrared spectroscopy (FTIR) was used for this purpose. The use of FTIR allows for the entire wavelength range between 1000nm and 5500nm to be studied with the detectors we have available. This range covers all the emitters of interest in a silicon on insulator platform.\cite{Wcenter,gcenter,Tcenter,selenium}. 

\begin{figure}[h]
\centering
\makebox[\textwidth]{\includegraphics[width=120mm]{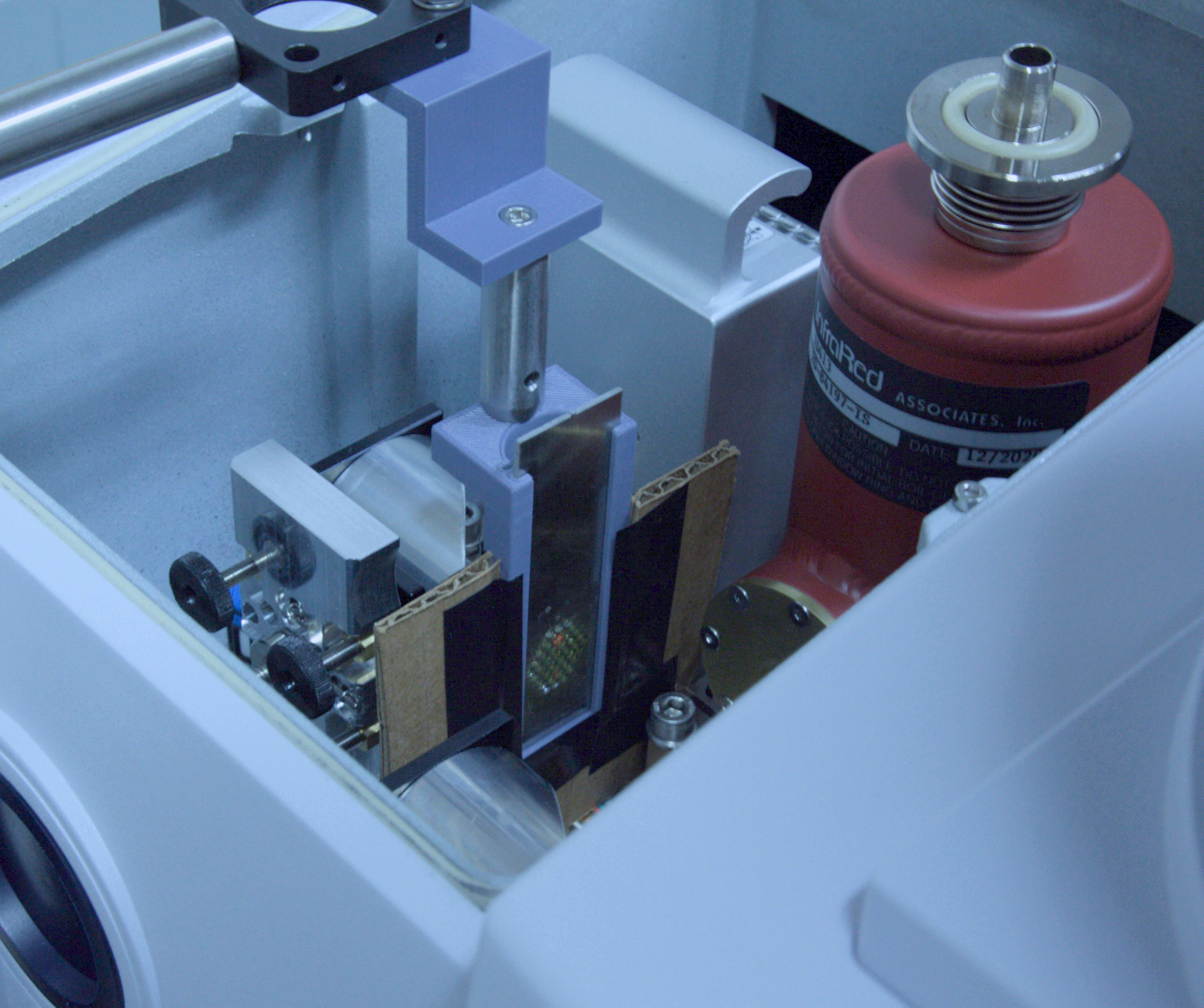}}
\caption{A resist sample mounted in the FTIR. The sample is mounted before the detector. A 3D printed holder allows for the samples to be quickly exchanged. Cardboard blocks stray light from the internal light source, maximizing the signal to noise ratio obtained.}
\label{fig:ftirsetup}
\end{figure}

Our Bruker Vertex 80 FTIR has been customized with a liquid nitrogen cooled germanium photodetector. This detector allows for the detection of extremely weak signals in the 800nm to 1800nm wavelength range. Unfortunately, this detector occupies the FTIR's sample compartment, making it impossible to place samples  in this location. As an alternative, the samples were placed in the detector compartment as shown in Fig. \ref{fig:ftirsetup}. 

\begin{figure}[h]
\centering
\makebox[\textwidth]{\includegraphics[width=120mm]{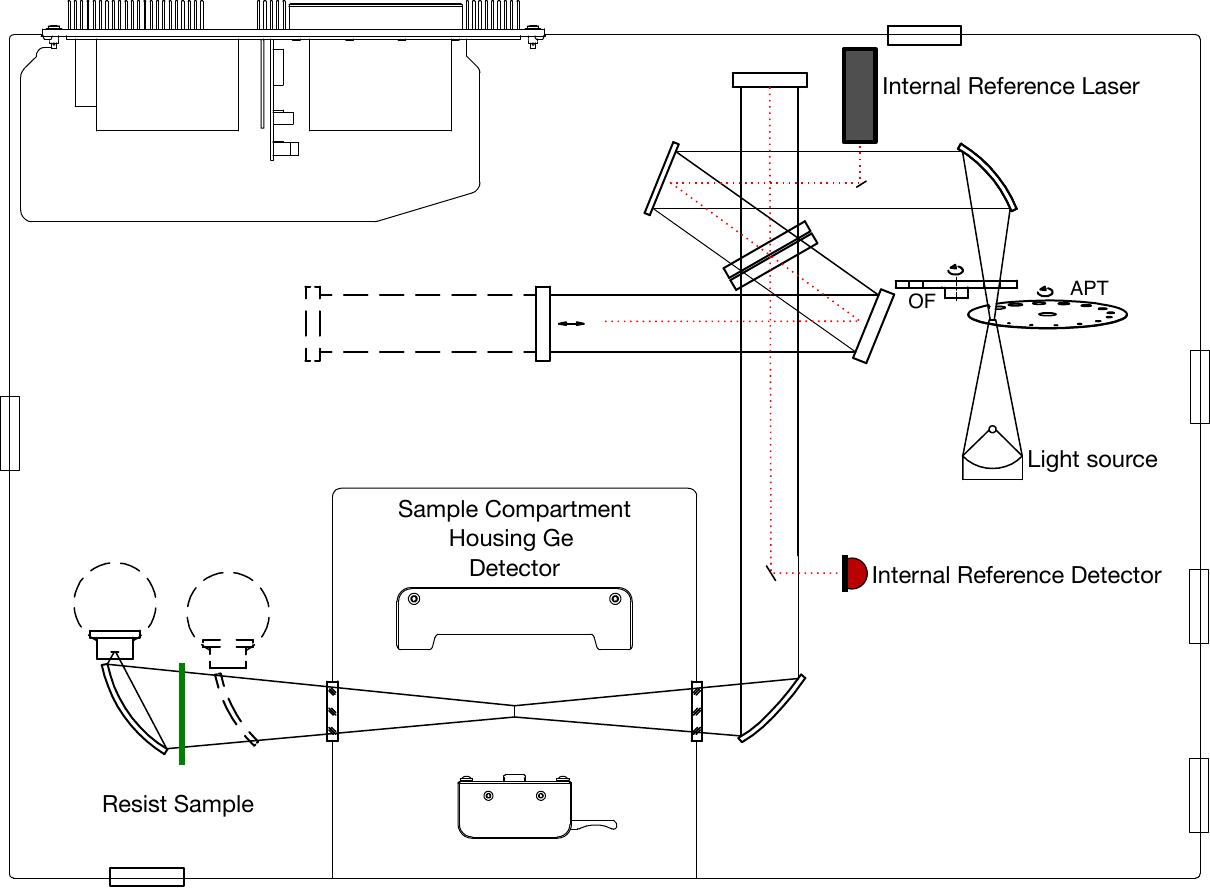}}
\caption{A simplified diagram of the optical path of the FTIR. Light from the internal source passes through the interferometer then the resist sample indicated by the green line before reaching the detector.}
\label{fig:ftirdiagram}
\end{figure}

A diagram of the optical path with this measurement configuration is shown in Fig. \ref{fig:ftirdiagram}. 

\begin{figure}[h]
\centering
\makebox[\textwidth]{\includegraphics[width=50mm]{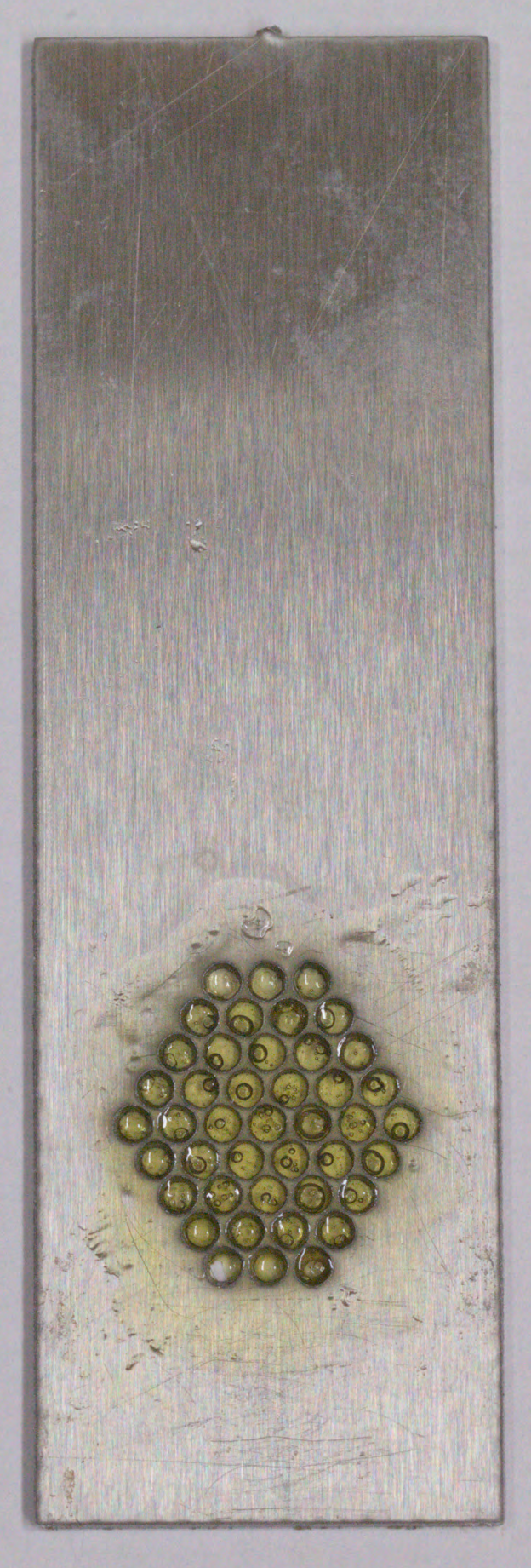}}
\caption{A photo of the mesh grid with a sample of photonic wirebond resist cured inside.}
\label{fig:meshgrid}
\end{figure}

To avoid any interference from the choice of substrate a sample of Vancore A resist was prepared on a metal grid by placing the resist on the grid and hardening it with the DynaCure DUV LED for 180 seconds on both sides. The metal grid can be seen in Fig. \ref{fig:meshgrid}.

\begin{figure}[h]
\centering
\makebox[\textwidth]{\includegraphics[width=120mm]{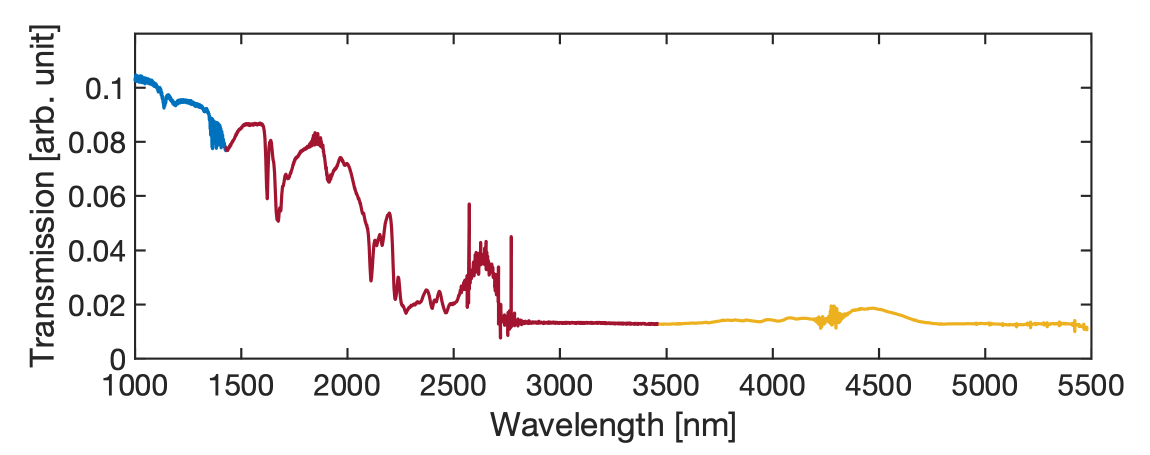}}
\caption{A normalized transmission spectrum for Vancore a prepared on a grid. The transmission values were scaled to match the spectrums between the different detectors and sources in their overlap regions. In blue is the InGaAs detector in combination with the NIR source. In red is the InSb detector in combination with NIR source. Finally in yellow is the InSb detector in combination with the MIR source. The sharp spikes present around 1400nm, 1800nm and 4500nm are from absorption caused by gases present in the atmosphere. The transmission through this sample is too low between 2800nm and 3600nm to see the absorption peaks from OH and CH bonds.}
\label{fig:FTIR}
\end{figure}

The high absorption for wavelengths longer then 2000nm means that PWB is not suitable for MIR defect centres such as selenium at 2902nm \cite{selenium}. The high absorption in the 2870 nm to 3435 nm region is from OH and CH bonds. It is unlikely this is able to be overcome using a different resist as  all organic photoresists will have these bonds \cite{ftirspectrums,ftirresist1,ftirresist2}.

\section{Assembling Process Development}
\subsection{Choice of Glue}
Samples that were assembled using vacuum compatible UV epoxy (Masterbond® UV10TKLO-2 \cite{UV24TKLOref}) did not survive cooldown. Although it is desirable to use UV epoxy that can cure within seconds using UV light during the assembly process, it is observed that even with a relatively slow cooling process (around 4 hours), this epoxy still suffers macroscopic cracking at low temperatures [Fig 4(d)]. This could be due to internal thermal stress within the epoxy and from shear or compression stress between the different materials. This damage made it prone to detach completely after being cooled. As a result, a cryo- and vacuum-rated two-part epoxy (Masterbond® EP29LPSPAO-1 Black \cite{EP29LPSPAOref}) is used instead for high mechanical and dimensional stability. The epoxy is also thermally conductive which helps the sample maintain good thermal contact with the sub-mount and consequently the cold stage. Additionally, reducing the amount of epoxy used and maintaining an even spread of epoxy on both sides of the FA and chip was found to help prevent mechanical failure. To ensure the sample remains aligned during the epoxy’s multi-hour curing process, small amounts of UV glue (Bondic® \cite{bondicref}) is first used to secure the aligned components onto the sub-mount. Once the epoxy has fully cured, this UV glue is easily and cleanly removed using a scalpel.

\subsection{Thermal Contraction Matching and Stress Management}
\label{thermalcontractionsection}
\begin{figure*}[h!]
\centering
\makebox[\textwidth]{\includegraphics[width=130mm]{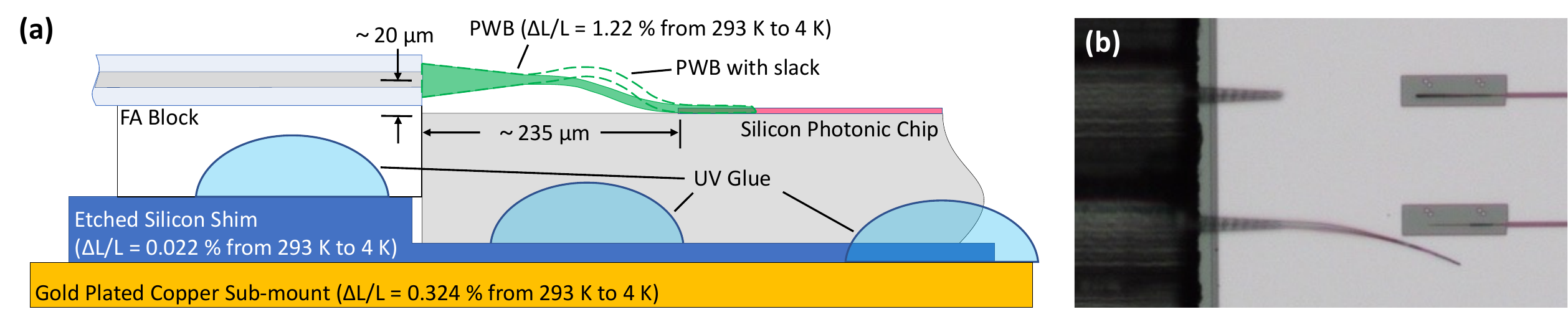}}
\caption{(a) Side schematic view of a failed sample that had multiple failure modes: 1. The sample used etched silicon shim to optimize the height. However, it has relatively small thermal contraction compared to the PWB polymer and copper sub-mount. 2. The FA block was pushed against the chip edge which did not leave room to compensate for the PWB polymer shrinkage. 3. The sample used UV glue that is prone to cracking at low temperature. This configuration is optimum for room temperature samples for easy assembly, reduced resist trapping within crevices, large bond bending radius, and reduced bond length. Slack is added to the bonds to help compensate for the shrinkage as demonstrated by the dashed bond, but all the bonds continue to break. (b) Optical image of broken bonds that have peeled off from the surface tapers.}
\label{fig:RTassembly}  
\end{figure*}

The cryogenic PWB assembly technique is modified from the room temperature technique to achieve successful operation down to cryogenic temperatures. Normally, at room temperature, it is ideal to aim for shorter photonic wire bonds with a minimal number of bends and minimum bending radius (> 80 \(\mu\)m) for low insertion loss (demonstrated by solid bond in Fig. \ref{fig:RTassembly}(a)). An etched silicon shim is used to raise the height of the FA such that the fiber cores are roughly 20 \(\mu\)m above the chip surface. The FA face is also butted against the chip edge to ease the assembly process and prevent any trapping of unexposed resist in the crevice between the FA and chip edge. However, this configuration is not suitable when the sample, which includes the silicon photonic chip and shim, copper sub-mount, and polymer-based PWB, undergoes thermal contraction during cooldown. Polymer experiences the highest amount of thermal contraction (\(\Delta\)L/L = 1.22 \%) from 293 K to 4 K followed by copper (\(\Delta\)L/L = 0.324 \%) and silicon (\(\Delta\)L/L = 0.022 \%) \cite{ekin2006experimental,reed1983materials}. Optical images of the failed samples show that PWBs with this configuration experience high tensile stress that resulted in the bonds being peeled off from the surface taper and thus failing. This suggests there is both a vertical and horizontal stress is present during cool down. Subsequently, a series bonds with slack added (demonstrated by the dashed bond in Fig. \ref{fig:RTassembly}(a)) was written to compensate for the thermal contraction. Here, slack is added to the bond by increasing the “height” of the middle bond section and the amount of slack is quantified by the percentage of length difference compared to a straight line. The slack was varied from 0.63 \% to 3.65 \%. However, the bonds continue to peel away and detach from the surface taper with no noticeable improvements even for the bonds with the highest amount of slack [Fig. \ref{fig:RTassembly}(b)].

\subsection{Choice of Sub-mount and Shim}
\label{shim_supplement}
As described in section \ref{thermalcontractionsection} the samples that utilize silicon shims between the copper sub-mount, the FA and the chip to achieve optimal height matching did not survive cooling. Both configurations, with and without a gap between the FA and the chip, experienced failed bonds. This suggests that silicon is not a suitable shim material as its relatively low coefficient of thermal expansion (CTE) does not match the high CTE of the polymer bonds. It is also not a good thermal conductor. As a result, the FA and chip is glued directly to the copper sub-mount and no shim is used.

\section{Experimental Setup}
\label{measurement_supplement}

The gold plated copper sub-mount of the sample is bolted to the gold-plated copper cold stage by two stainless steel screws. Any excess fiber is carefully wrapped and secured to the radiation shield using Kapton\texttrademark{} tape and separated from touching the wall of the outer vacuum chamber. Any bending of the fibers is minimized and the targeted minimum bend radius is larger then 30 mm.  

Coiling and anchoring the excess fiber to the radiation shield can create torsion and bending that introduces unwanted birefringence \cite{ulrich1980bending,ulrich1979polarization}. In addition, the fiber is subject to refractive index change during cooling as the radiation shield reaches down to 40 K. This can induce polarization dependent loss in the measurement as the polarization of the laser light becomes random when travelled through the fiber and which causes a mode mismatch between the PWB and the polarization sensitive surface taper. Although a polarization controller can be used to optimize the polarization during the whole process, it is limited to only one wavelength. As a result, a polarization scrambler is used to minimize the degree of polarization across the measuring spectrum such that the random polarization dependent loss is decoupled from the insertion loss (IL) calculation of the PWB. The detector’s averaging time is set to 2 ms (500 Hz), which is higher than the scrambler’s recommended cut-off frequency of 10 kHz, to achieve a low degree of polarization of < 5 \%. The laser was sweept at a speed of 5nm/s between 1490-1640nm. Measurement of multiple samples using a polarization paddle controller optimized for TE coupling versus a scrambler show a consistent excess measurement loss of 0.7 dB ± 0.2 dB per PWB when measured using the depolarized method. 

\section{Extraction of PWB Insertion Loss}

\begin{figure}[h!]
\centering
\makebox[\textwidth]{\includegraphics[width=120mm]{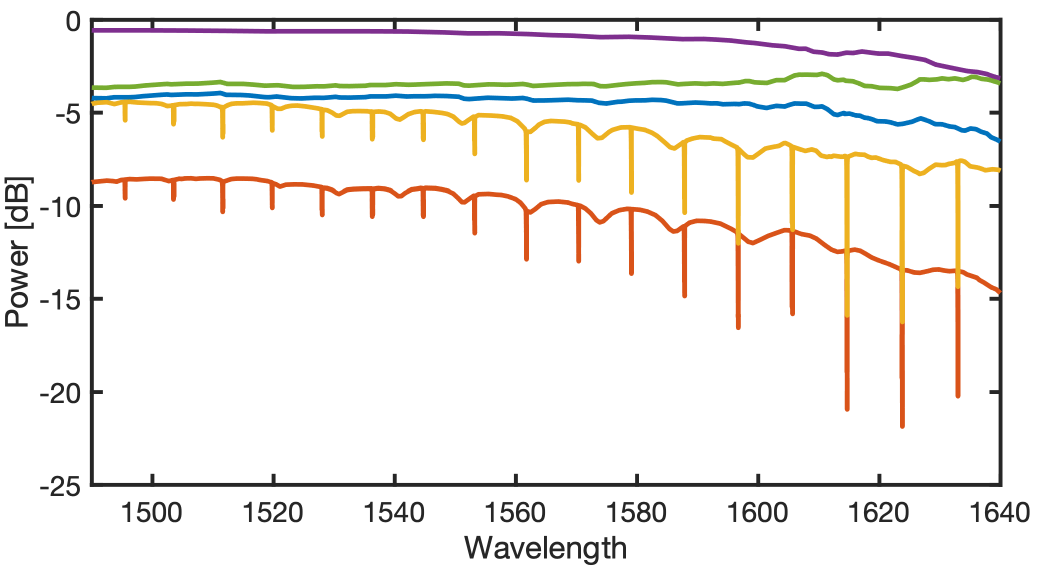}}
\caption{Plot showing the sources of loss that was calibrated out of the measurement to extracted the transmission through only the two PWBs and ring resonator.}
\label{fig:Cali}
\end{figure}

Fig. \ref{fig:Cali} shows the measured transmission through the whole optical pathway (red), the loss from the scrambler and fibers (purple) and UHV feedthrough (green), and the calibrated transmission through only the sample which consists of the two PWBs and ring device (yellow). The data shown in blue is the loss through the scrambler, fibers, and UHV feedthrough (purple + green).

\section{Multiple Cool Down Measurement}

\begin{figure}[h!]
\centering
\makebox[\textwidth]{\includegraphics[width=120mm]{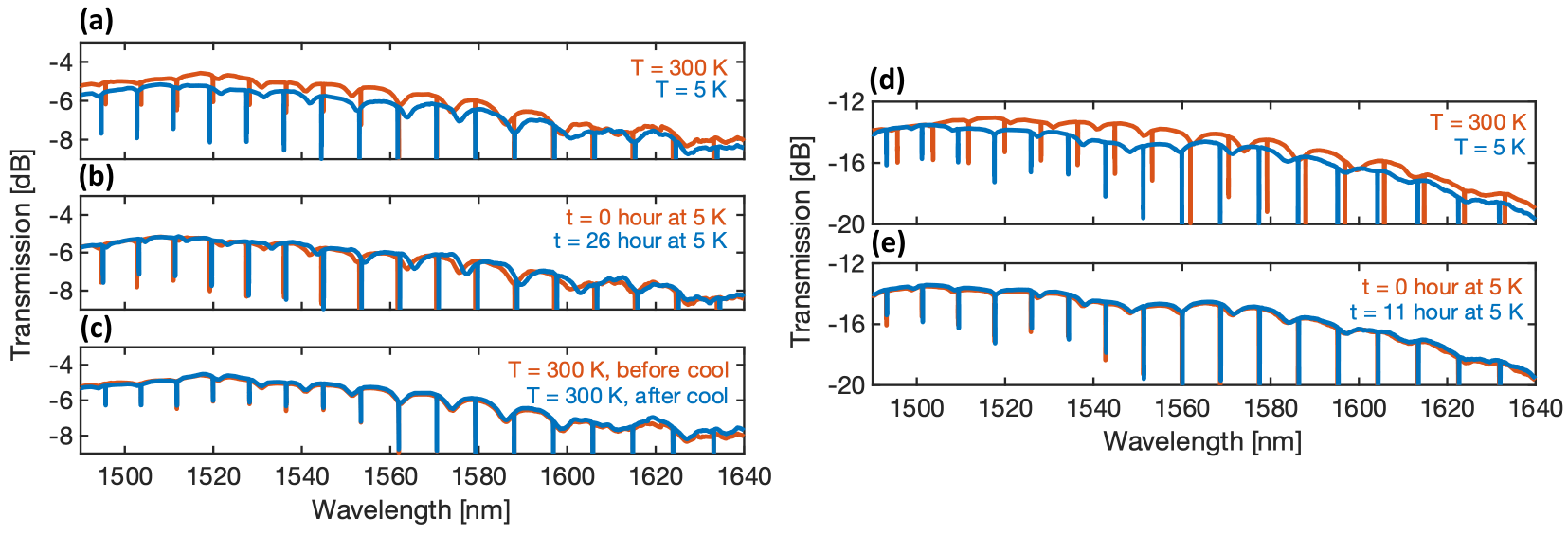}}
\caption{Plot (a)-(c) shows the calibrated transmission through the two PWBs and ring resonator device for the second cool down cycle. Plot (d) and (e) shows the uncalibrated transimssion through the whole optical path for the fourth cool down cycle.}
\label{fig:Repeatability}
\end{figure}

The micro ring resonator shown in Fig. 5(a) of the main paper was cooled several times to test the repeatability of the PWBs. Fig. \ref{fig:Repeatability} (a)-(c) shows the calibrated measurement through the two PWBs and ring resonator for the second cool down following the first cool down shown in Fig. 5 (b)-(d). We show the fourth cool down in Fig. \ref{fig:Repeatability} (d)-(e) (another device was measured for the third cool down). There is no data for warm up because the bond broke during high power testing. The data is not calibrated and the extra loss compared to Fig. \ref{fig:Cali} is from added attenuation for the high power testing. The measurement shows the transmission of the sample slightly degrades from 300 K to 5 K, remains the same for many hours at 4 K, and returns back to its original value after warming up. 

\section{Ring Resonator in Non-UHV Cryostat}

\begin{figure}[h!]
\centering
\makebox[\textwidth]{\includegraphics[width=120mm]{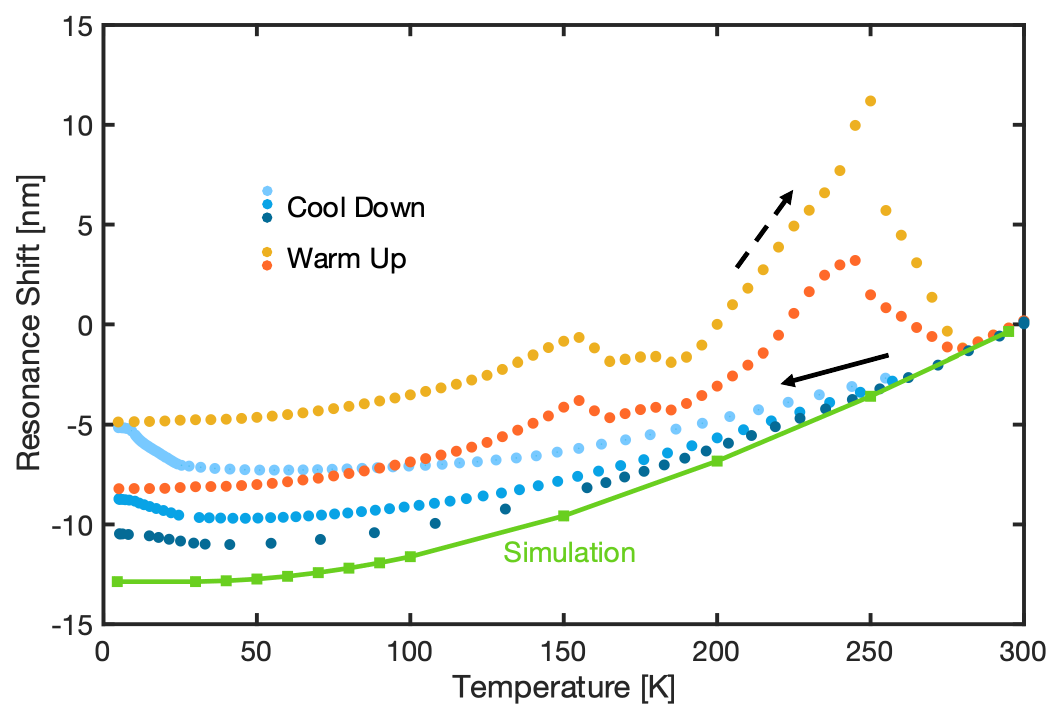}}
\caption{Plot showing the shift of the resonance wavelength around 1550 nm as a function of temperature without baking. The resonances in blue are measured during a cool down cycle while the resonances in yellow are measured during a warm up cycle.}
\label{fig:RingNoUHV}
\end{figure}

Fig. \ref{fig:RingNoUHV} shows the measured resonance shift of a ring resonator inside an unbaked cryogenic chamber with pressure of \(\sim 9 \times 10^{-5}\)  mbar  over several cooling and warming cycles compared to the expected theoretical resonance shift assuming in perfect vacuum. We expect a 12.5 nm blue shift around 1550 nm (green) due to the decrease in effective refractive index (\(n_{eff}\)) of the waveguide. However, the first cool down measurement (black arrow) shows a resonance shift of only 5.7 nm (light blue) which suggests there is a counter effect on the \(n_{eff}\) of the waveguide during cooldown. This is likely due to the deposition of gas molecules that are inside the chamber, such as water and nitrogen, onto the surface of the chip. Moreover, the resonance undergoes a steep red shift (dashed arrow) during warm up starting around 185 K which is likely due to the deposition of water molecules that evaporated from the warmed chamber on to the still cold sample. Water has a vapour pressure of \(\sim 9.4 \times 10^{-5}\)  mbar at 183 K \cite{buck1981new}. Once all gas molecules have evaporated off the sample at 300K, the resonance returns to the same wavelength prior to cooling. The resonance shift matches closer to the simulation with each new cool down cycle as gas particles are removed from the chamber by the turbo pump over time.

\section{Modeling of Ring Resonator Cool Down}

The change of refractive index of silicon as a function of temperature is the main effect in causing the resonance shift of the ring resonator during cool down. The \(m^{th}\) order resonance wavelength of a given ring with radius R is 

\begin{equation}
\frac{m\lambda_m}{n_{eff}} = 2\pi R \label{eq:eq1}
\end{equation}

\(n_{eff}\) is the effective index of the waveguide mode and is calculated using Lumerical MODE. For a fixed bent waveguide at temperature, T, the \(n_{eff}\) can be fitted to a Taylor expansion approximation:

\begin{equation}
n_{eff}(T,\lambda) = n_1(T) + n_2 (T)(\lambda - \lambda_o) + n_3(T)(\lambda - \lambda_o)^2 
\label{eq:eq2}
\end{equation}

Using Eq. \ref{eq:eq1} and Eq. \ref{eq:eq2}, the resonance wavelength of the \(m^{th}\) order is derived. 

\begin{equation}
\lambda_m(T) = \frac{2\pi R}{m}[n_1(T) + n_2(T)\lambda_m + n_3(T)\lambda_m^2] \label{eq:eq3}
\end{equation}

The waveguide height and width in the MODE simulation is adjusted to match the fabricated dimension measured by the AFM and SEM. The width is finely tuned until the FSR and resonance wavelengths match close to the measured transmission spectrum. To model \(\lambda_m\) at different temperatures, the refractive index data from \cite{frey2006temperature,leviton2006temperature,komma2012thermo} at 1500 nm is fed into the MODE simulation as a perturbation to the refractive index of silicon and silicon dioxide to extract the new \(n_{eff}(T,\lambda)\). Here, the effect due to thermal contraction is neglected as it is a 100 factor smaller than the thermo-optic effect \cite{you2020chip}. Additionally, the refractive index perturbation is assumed to be the same for the range of the wavelength shift and for resonances near 1500 ± 50 nm . Fig. 6(a) in the main paper shows the resonance wavelength versus temperature over multiple cool down and warm up cycles and the data shows good agreement to the model. 

\section{Power Limit}
\begin{figure}[h!]
\centering
\makebox[\textwidth]{\includegraphics[width=120mm]{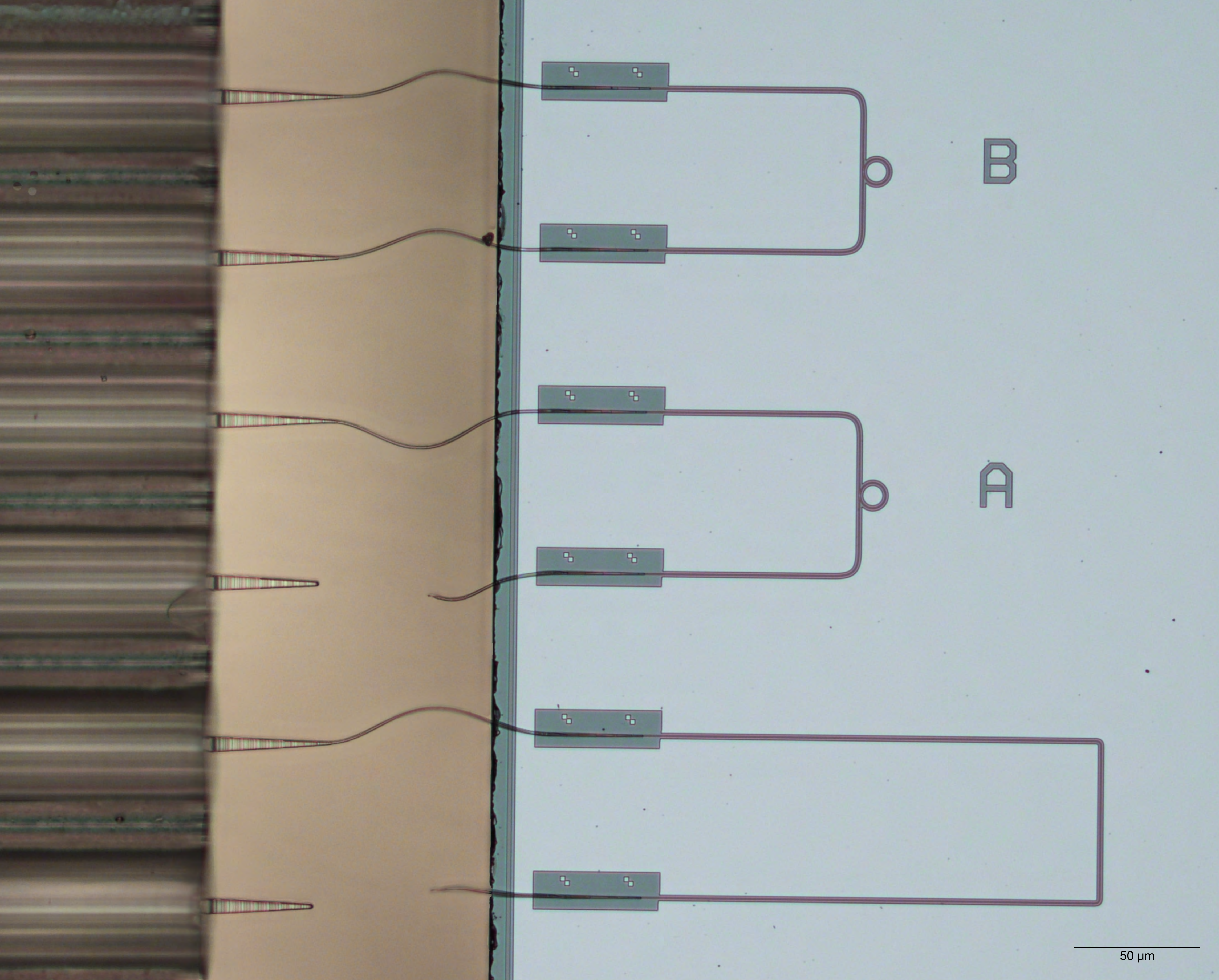}}
\caption{A focus stacked optical microscope image of the failed photonic wire bonds. One can see the evaporated bond from the high power testing in both the ring resonator structure labeled A and the waveguide loop back below A. The ring resonator structure labeled B is present with two intact bonds; this device was not powered during the experiment and is present as a reference.}
\label{fig:burnbond}
\end{figure}

An optical microscope image of the photonic wirebonds that evaporated during high power testing is shown in Fig. \ref{fig:burnbond}. Based on the evaporated section visible on the bottom waveguide loop back and the MRR labled A it was conclude that the lack of convection cooling in the vacuum environment was responsible for the much lower power limit.

\bibliography{osa-supplemental-document-template}